\documentclass[prl,twocolumn,showpacs,superscriptaddress,nofootinbib]{revtex4-2}
\pdfoutput=1
\usepackage{graphicx}
\usepackage{epstopdf} 
\usepackage{amsmath}
\usepackage{amssymb}
\usepackage{mathrsfs}
\usepackage{bm}
\usepackage{latexsym}
\usepackage{color}
\usepackage[usenames,dvipsnames]{xcolor} 
\usepackage[colorlinks=true,citecolor=MidnightBlue,linkcolor=RubineRed,urlcolor=MidnightBlue]{hyperref} 
\usepackage{float} 
\usepackage{multirow}         

\def\x{{\rm\bf x}}

\def\la{\langle}
\def\ra{\rangle}

\newcommand{\beq}{\begin{equation}}
\newcommand{\eeq}{\end{equation}}
\newcommand{\beqa}{\begin{eqnarray}}
\newcommand{\eeqa}{\end{eqnarray}}

\usepackage{tikz,xcolor,hyperref}
\definecolor{lime}{HTML}{A6CE39}
\DeclareRobustCommand{\orcidicon}{
\begin{tikzpicture}
\draw[lime, fill=lime] (0,0)
circle[radius=0.16]
node[white]{{\fontfamily{qag}\selectfont \tiny \.{I}D}}; 
\end{tikzpicture}
\hspace{-2mm}
}
\foreach \x in {A, ..., Z}{%
\expandafter\xdef\csname orcid\x\endcsname{\noexpand\href{https://orcid.org/\csname orcidauthor\x\endcsname}{\noexpand\orcidicon}}
}

\begin{document}
\title{Evolution of  Vortex Strings after a Thermal Quench in a Holographic Superfluid}

\author{Chuan-Yin Xia}
\affiliation{Center for Theoretical Physics, Hainan University, Haikou 570228, China}

\author{Andr\'as Grabarits\orcidA{}}
\affiliation{Department  of  Physics  and  Materials  Science,  University  of  Luxembourg,  L-1511  Luxembourg, Luxembourg}

\author{Hua-Bi Zeng \orcidB{}}
\email{zenghuabi@hainanu.edu.cn}
\affiliation{Center for Theoretical Physics, Hainan University, Haikou 570228, China}

\author{Adolfo del Campo \orcidC{}}    
\email{adolfo.delcampo@uni.lu}
\affiliation{Department  of  Physics  and  Materials  Science,  University  of  Luxembourg,  L-1511  Luxembourg, Luxembourg}
\affiliation{Donostia International Physics Center,  E-20018 San Sebasti\'an, Spain}

\begin{abstract} 
The formation of topological defects during continuous phase transitions exhibits nonequilibrium universality. While the Kibble-Zurek mechanism (KZM) predicts universal scaling of point-like defect numbers under slow driving, the statistical properties of extended defects remain largely unexplored across both slow and fast protocols. We investigate vortex string formation in a three-dimensional holographic superfluid. For slow quenches, the vortex string number follows KZM scaling, while for rapid quenches, it exhibits complementary universal scaling governed by the final temperature. Beyond the vortex string number, the loop-length distribution reveals a richer structure: individual loops follow the first-return statistics of three-dimensional random walks, $P(\ell) \sim \ell^{-5/2}$. While the total vortex length distribution remains Gaussian, its cumulants obey universal scaling laws with varying power-law exponents, and thus differ markedly from those observed in point-defect systems, indicating distinct statistical features of extended topological defects.
\end{abstract}

\maketitle
Characterizing universal physics far from equilibrium is an enduring challenge. The Kibble-Zurek Mechanism (KZM) offers a rare bridge between equilibrium phenomena and dynamics, estimating the density of defects that spontaneously form during the dynamics of a continuous phase transition by using critical scaling theory \cite{Kibble1976,Kibble1980,Zurek1985,Zurek1996,Dziarmaga05,Polkovnikov2011,delCampo2014}.
When the control parameter $\lambda(t)$ driving the phase transition is varied across its critical value $\lambda_c$, both the equilibrium correlation length $\xi=\xi_0/|\varepsilon|^{\nu}$ and the relaxation time $\tau=\tau_0/|\varepsilon|^{z\nu}$  exhibit a universal power-law behavior as a function of $\varepsilon=(\lambda-\lambda_c)/\lambda_c$, governed by the critical exponents $\nu$ and $z$. Critical slowing down makes the dynamics nonadiabatic, leading to the formation of topological defects. The system response to the quench occurs on a time scale known as the freeze-out time, which for a linear ramp $\lambda(t)=\lambda_c(1-t/\tau_Q)$, takes the form  $\hat t \propto \tau_Q^{z\nu/(1+z\nu)}$.  KZM estimates the effective nonequilibrium correlation length $\hat{\xi}$ by the equilibrium value at the freeze-out time, i.e.,  $\hat{\xi}=\xi[\varepsilon(\hat{t}]$. In $D$-spatial dimensions, a system of size $L\gg\hat\xi$ is partitioned into a number of causally disconnected regions or protodomains $N_d=(L/\hat\xi)^D$, which determines the average defect number $\langle N\rangle$,
\begin{equation}
\langle N\rangle \propto \hat\xi^{-D} \propto \tau_Q^{-D\nu/(1+z\nu)}.
\label{kzlaw}
\end{equation}
This scaling law assumes point-like defects such as kinks and vortices. 
Numerical tests of the KZM in one~\cite{Zurek1997,delCampo10}, two~\cite{Zurek1998}, and three dimensions \cite{Zurek1999}  have confirmed the nonequilibrium scaling laws predicted by KZM using Landau–Ginzburg dynamics with mean-field critical exponents $\nu=1/2$ and $z=2$. However, in scenarios of 3$D$ $U(1)$ symmetry breaking characterized by the formation of spatially extended vortex strings, an intriguing distinction appears: while the number of vortex-line elements is expected to scale as $\tau_Q^{-3/4}$, the total string length $\langle L\rangle$ follows a weaker scaling \cite{Zurek1999}, identical to the KZM scaling for the vortex number in  2$D$ \cite{Zurek1998,Zurek1999}. This observation clearly highlights that extended defects in 3$D$ exhibit fundamentally richer and distinct statistical features than point-like defects.

A wide variety of experiments in $D=1$ \cite{Digal1999,Carmi2000,Monaco2002,Monaco2003,Monaco2006,Pyka2013,Ulm2013,Ejtemaee2013,Bando2020,King2022,PRXQuantum.5.040320,keesling2019quantum} and $D=2$ \cite{Maniv2003,Golubchik2010,Sadler2006,Weiler2008,Lamporesi2013,Navon2015,Donadello2016,Ko2019,manovitz2025quantum,lee2024universal} have established the robustness of KZM predictions in lower dimensions.
 Vortex string formation in 3$D$, as originally envisioned by Kibble \cite{Kibble1976,Kibble1980}, has been studied in superfluid helium \cite{Zurek1985,Zurek1996,Dziarmaga05,Polkovnikov2011,delCampo2014}, liquid crystals, helium experiments \cite{Chuang1991,Bowick1994,Hendry1994,Bauerle1996,Ruutu1996,Dodd1998}, and multiferroics \cite{lin2014topological}. The formation of 3$D$ Ising domains via the KZM has recently been reported \cite{2023Du}. For spatially extended defects with defect dimension $d$ (e.g., $d=1$ for a vortex string), the KZM needs to be modified.  A general expectation is that the defect density scales as \cite{Antunes1999,delCampo2014,hartnoll2018}
\begin{equation} \label{kzlawL}
\langle N\rangle \propto \tau_Q^{-(D-d)\nu/(1+z\nu)}.
\end{equation}
However, this estimate is based on dimensional analysis, and the meaning and definition of defect density have received little attention, even in the scarce literature that tests it \cite{Antunes1998}. For instance, for vortex strings, one may distinguish between the loop number and the string length.

Beyond the slow-quench regime, KZM exhibits a universal breakdown as a function of the final quench depth $\epsilon_f=(\lambda_f-\lambda_c)/\lambda_c$ set by $\tau\propto\epsilon^{-z\nu-1}_f$~\cite{Zeng2023,Chesler2015}. For point-like defects, the average defect number reaches a fast-quench plateau, as also demonstrated in various experiments~\cite{Donadello2016,Ko2019,Goo2021,Goo2022}. Its height and the corresponding freeze-out timescale inherit a universal power-law scaling dictated by the quench depth rather than the driving rate
\begin{equation} \label{rapidn}
\langle N \rangle \propto \epsilon_f^{D\nu},\qquad \hat t\propto\epsilon^{-z\nu}_f.
\end{equation}
While these predictions have been justified in a few showcase scenarios in 1$D$~\cite {Wang:2025swz,Rao:2025xku,Xia2023str,PhysRevB.110.064317} and 2$D$ systems~\cite{xia2024disk}, no works have established their validity for spatially extended defects. 

In addition to these scaling laws for the mean value $\kappa_1=\langle N\rangle$, the variance $\kappa_2=\langle N^2\rangle-\langle N\rangle^2$, and higher order cumulants have been shown to be universal and follow the same power-law as the mean, implying the universal character of the full number distribution of point-like defects. The universality of the defect statistics beyond the KZM regime has been proposed for slow quenches~\cite{delCampo2018,GomezRuiz2020}, and is supported by experiments~\cite{Cui2020,Goo2021,Bando2020,King2022} and simulations~\cite{Xia2020,Mayo2021,delCampo2021,Subires2022,li2022holographic,PhysRevResearch.4.023201}, while limited findings are available for fast quenches in 1$D$~\cite{Wang:2025swz,Rao:2025xku} and 2$D$ ~\cite{xia2024disk}. 
As a result, the statistical features of extended defects such as vortex strings remain to be elucidated.

In this Letter, we establish the universality of spatially-extended defect generation by studying vortex-string formation in a three-dimensional superfluid. Within this framework, we establish universal scaling laws for the average number and length of vortex strings, along with their corresponding variances, in both slow and fast quenches.  Their fluctuations approach those of a Gaussian distribution, but most strikingly, the variance of the vortex-line length shows a scaling different from the mean in both slow and fast quenches, revealing fundamentally distinct statistical behavior from that in previous studies.

We benchmark our predictions using the AdS/CFT correspondence \cite{Maldacena1999,Gubser1998,witten1998}, which relates a system in $D+1$ spacetime dimensions to a $D+2$ gravitational problem, and provides a powerful approach to explore non-equilibrium phenomena in strongly correlated condensed matter systems (AdS/CMT) \cite{zaanen2015,ammon2015,hartnoll2018,zaanen2021,Liu2020}. 
Specifically, we employ a holographic superfluid model to study vortex-string formation, where the universal equilibrium properties emerge from a charged scalar in an AdS black hole background, tuned via the black hole temperature \cite{Gubser2008,Hartnoll2008,2009holosuperfluid}. This framework has successfully captured far-from-equilibrium KZM dynamics in one and two spatial dimensions \cite{Xia2020,delCampo2021,li2022holographic,PhysRevResearch.4.023201,Xia2023str,xia2024disk,Chesler2015,sonner2015universal,Natsuume2017,Zeng2019,delCampo2021,Yang:2025bsw}, and has been used to simulate quantum vortex dynamics in 2$D$ \cite{Hong2014,du2015,2023PhRvL.131v1602L,2021PhRvL.127j1601W,Lan:2016cgl,2023PhRvB.107n4511Y,2023PhRvB.107n4511Y,Yang2024hom,Su2023,xia2022Abrikosov,Lixin2020,Xia:2019eje} and vortex-string dynamics of quantum turbulence in 3$D$ \cite{Zeng:2024rwn,Wittmer:2024ifm}.

\begin{figure}
    \centering
    \includegraphics[width=0.8\linewidth]{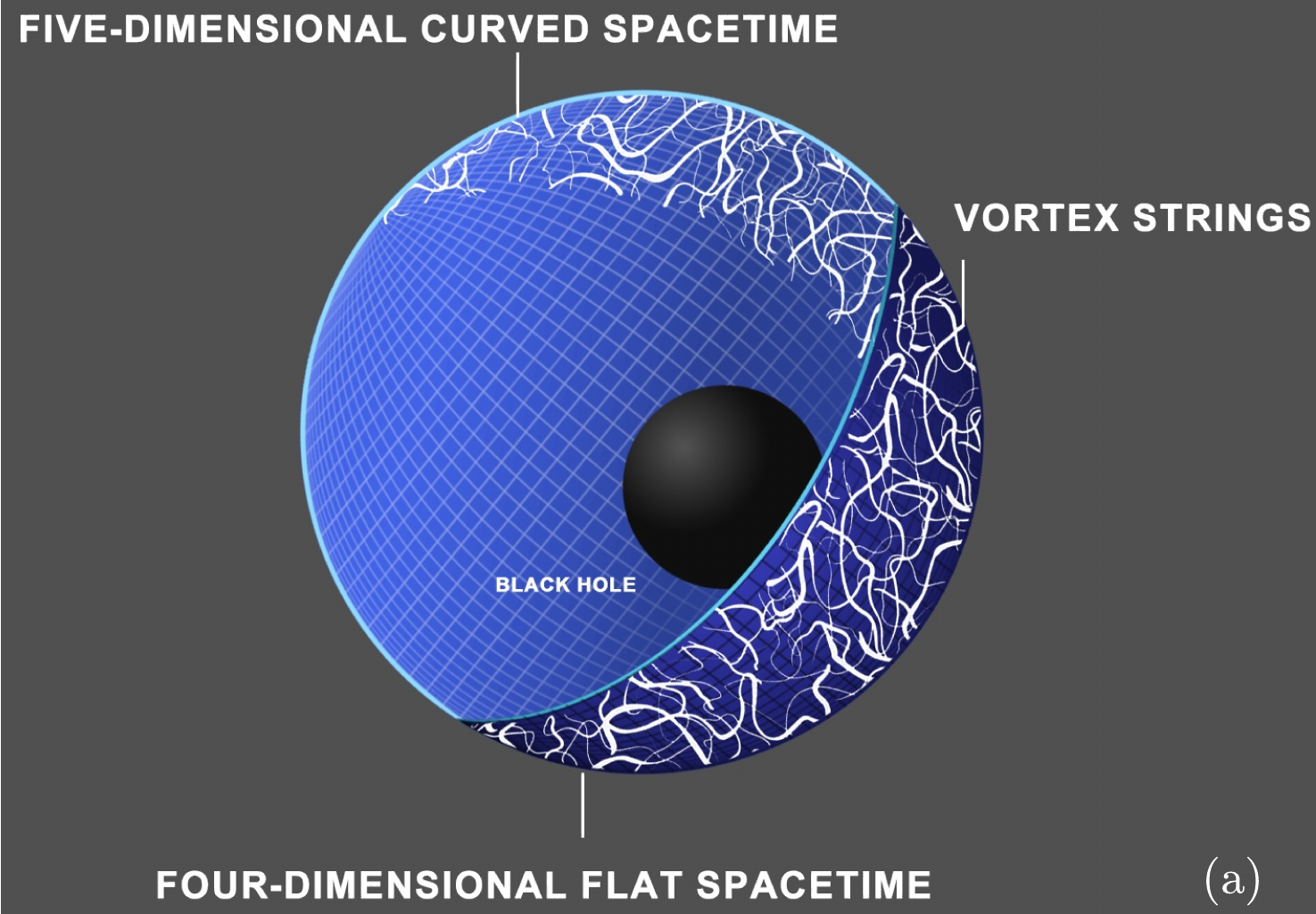}
\includegraphics[width=0.8\linewidth]{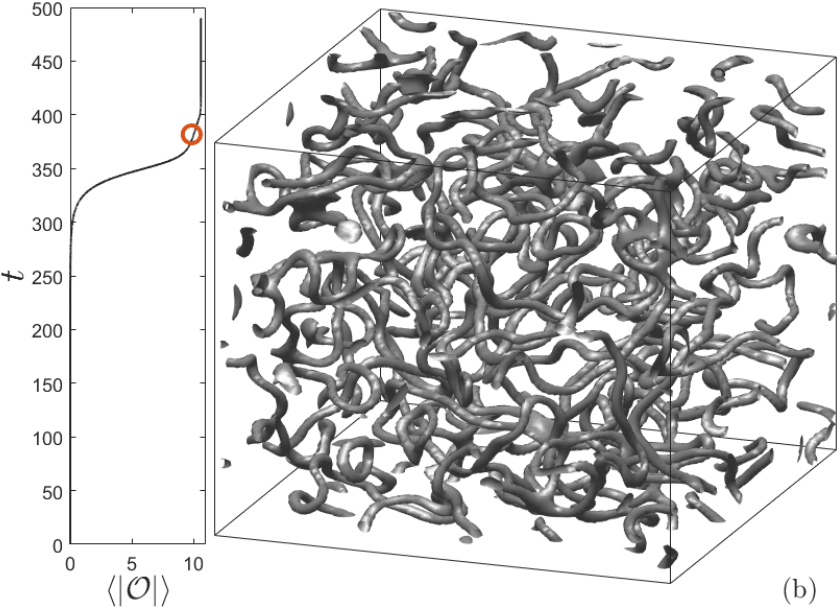}
    \caption{(a) Schematic representation of the holographic mapping between a gravity theory with a five-dimensional black hole and a strongly-coupled thermal field theory supporting vortex strings. (b) Dynamics of the order parameter across the critical point and vortex string formation. The left panel displays the time evolution of the average amplitude of the order parameter, while the right panel shows an isosurface of the order parameter at the instant indicated in the red ring. 
     The size of this holographic superfluid is 100×100×100, with its final temperature and quench time recorded as  $T_f=0.8T_c$ and $\tau_Q=2000$  respectively.}
    \label{fig1}
\end{figure}

The holographic model captures the nonequilibrium phase-transition dynamics of a $U(1)$ theory through a fully dynamical gravitational evolution, providing a first-principle description of temperature quenches across the critical point. We focus on the nonequilibrium $U(1)$ symmetry-breaking transition in three spatial dimensions, using the action \cite{Gubser2008,Hartnoll2008,2009holosuperfluid}
\begin{equation}\label{action}
S\!=\!\int\!d^{D+2}x\sqrt{-g}\Big[R+ \frac{\Lambda}{L^2}+\frac{1}{q^2}\big(-\tfrac14 F^2-( |D\Psi|^2-m^2|\Psi|^2)\big)\Big],
\end{equation}
where $|\Psi|$ is the charged scalar field, $F^2:=F^{MN}F_{MN}$, the electromagnetic field $F_{MN}=\partial_M A_N-\partial_N A_M$, and $D_M:=\partial_M-i A_M$. 

In the probe limit, the matter fields do not backreact on the geometry. For $D=3$, solving only the gravity part of the action (\ref{action}) yields the background geometry of an AdS$_5$ Schwarzschild black hole
\begin{equation}
ds^2=\frac{\ell^2}{u^2}(-f(u) dt^2 -2Dtdu+ dx^2+dy^2+dz^2),
\end{equation}
with $f(u)=1-(u/u_h)^4$, where the extra bulk dimension $u$ roughly encodes the RG scale of the dual field theory.
The Hawking temperature $T=(\pi u_h)^{-1}$ sets the energy scale of the $3+1$-dimensional superfluid. Below the critical value $T_c$, a condensate of $\Psi$ forms in the bulk, spontaneously breaking the $U(1)$ symmetry and driving a second-order phase transition in the boundary theory. The resulting nonequilibrium critical dynamics generate vortex-string defects, classified by the homotopy group $\pi_1(S^1)\cong \mathbb{Z}$ and defect dimension $d=1$, in contrast to the point-like defects ($d=0$) realized in lower-dimensional systems.

We impose a linear cooling protocol,
\begin{eqnarray}
T(t)=
\begin{cases}
T_c(1-t/\tau_Q)  & 0 \leq t<t_f \\
T_f              & t \geq t_f
\end{cases},
\end{eqnarray} 
where $\tau_Q$ is the quench time of linear quenches. The end time of the quench $t_f$ is defined by setting $T(t_f)=T_f$
and is given by $t_f=\tau_Q(1-T_f/T_c)$.

By solving the full $4+1$-dimensional nonlinear PDEs of the holographic model for a boundary system of size $100^3$ as detailed in \cite{Supplementary}, we track the vortex-string dynamics during a thermal quench. A typical realization with $\tau_Q=2000$ and $T_f=0.8T_c$ is shown in Fig.~\ref{fig1}, displaying the vortex-string isosurface and the evolution of the order parameter $\langle|\mathcal{O}|\rangle$. Once $\langle|\mathcal{O}|\rangle$ reaches about 10\% of its final equilibrium value, it enters a rapid-growth regime that sharply defines the freeze-out time $\hat t$, at which we count the vortex strings.

The statistics of the vortex-line number in 3D superfluids exhibit richer features than for point-like defects in lower dimensions. Extended vortex strings form at high density in the early stages of the critical dynamics. Helmholtz’s theorem~\cite{Helmholtz1858} ensures that the flux of vortex strings is conserved, enforcing the formation of closed loops under periodic boundary conditions. Within this framework, the number of candidate sites for vortex-loop formation is determined by the KZM in a given volume, $\mathcal N = \mathrm{Vol}/(f\hat\xi^{3})$, with the mean field critical exponents $z=2$ and $\nu=1/2$ for the holographic superfluid transition, and $f$ being a fudge factor. Loop-formation events are assumed to occur independently with a uniform success probability $p$.
Thus, the loop number statistics follows a binomial distribution,
\begin{eqnarray}\label{eq: Binom}
    P(N)=\binom{\mathcal N}{N}p^N(1-p)^{\mathcal N-N}.
\end{eqnarray}
As a result, all loop-number cumulants are proportional to the mean and thus scale universally with the driving time or the quench depth,
\begin{eqnarray}
    \kappa_n(N)\propto \tau^{-3/4}_Q,\quad\kappa_n(N)\propto \epsilon^{3/2}_f,
    \label{knscalings}
\end{eqnarray}
for the slow and fast driving limits, respectively.

\begin{figure}[t]
    \centering
        \includegraphics[width=\linewidth,trim={0 4.2cm 0 12cm}, clip]{ 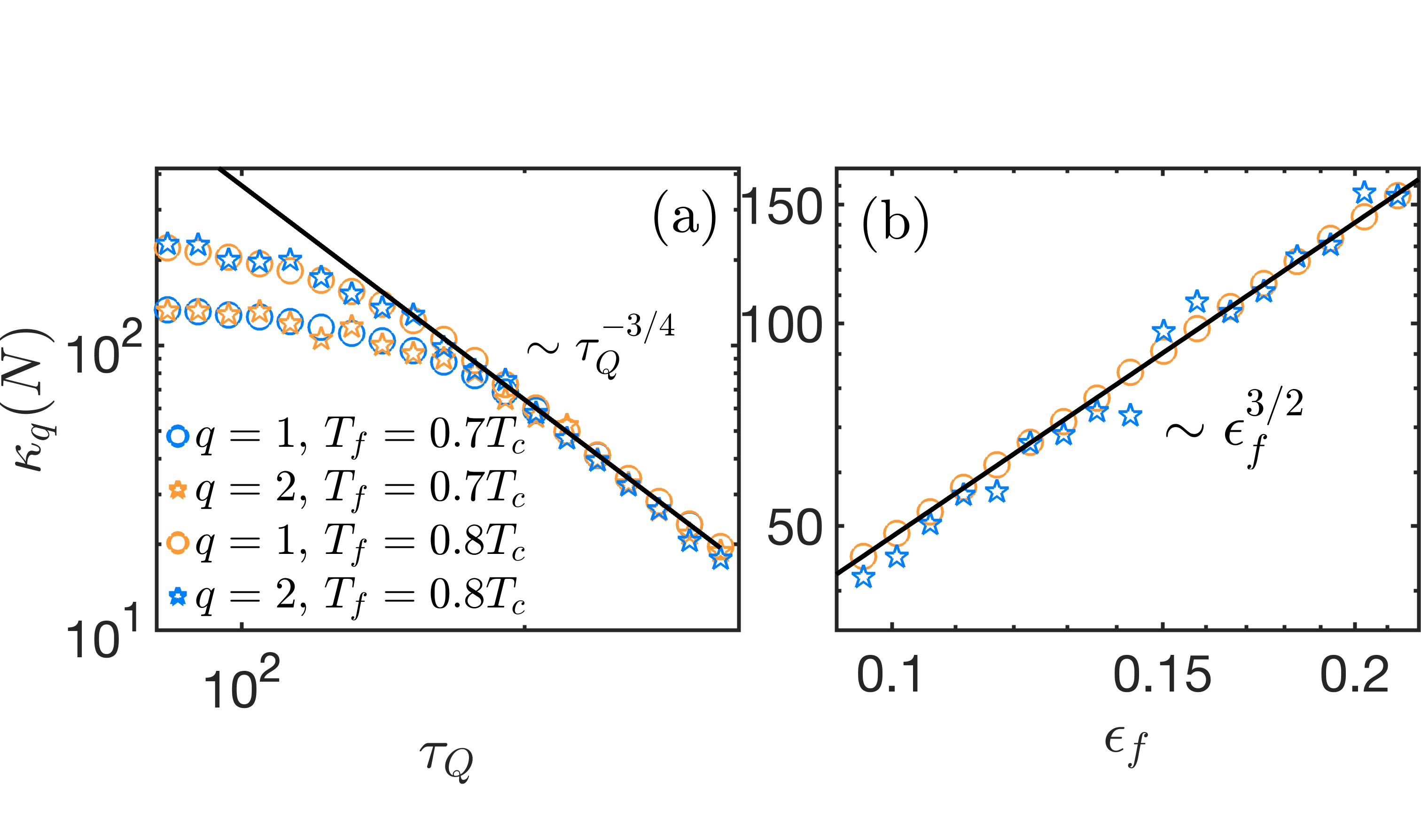}
    \caption{
        (a) Scaling of the first two cumulants of the vortex loop number distribution, exhibiting approximately the same power-law behavior for the two final temperatures, $T_f=0.7T_c,\,0.8T_c$. (b) Universal scaling of the first two cumulants at fast quenches ($\tau_Q=1$). Both regimes are well described by the power laws in \eqref{knscalings}.}
    \label{fig2}
\end{figure}

As shown in Fig.~\ref{fig2}(a), for slow quenches, the first two cumulants, obtained by averaging over $1000$ random realizations, are best fitted by $\kappa_{1,2} = (11700 \pm 500), \tau_Q^{-0.753 \pm 0.006}$, in agreement with theoretical predictions for driving times $\tau_Q \gtrsim 100$. The results are presented for two final temperatures, $T_f = 0.7T_c$ and $0.8T_c$, indicating the robustness of the scaling. For shorter quenches, $\tau_Q \lesssim 100$, the curves saturate to a plateau, indicating the breakdown of KZM scaling in the fast-quench regime. In this regime, the data also follow the expected scaling behavior, as shown in Fig.~\ref{fig2}(b), with the best fits $\kappa_{1,2} \approx (1710 \pm 30) \epsilon_f^{1.55 \pm 0.02}$.

\begin{figure}
    \centering
    \includegraphics[width=0.99\linewidth,trim={0 6.5cm 0 12cm},clip]{ 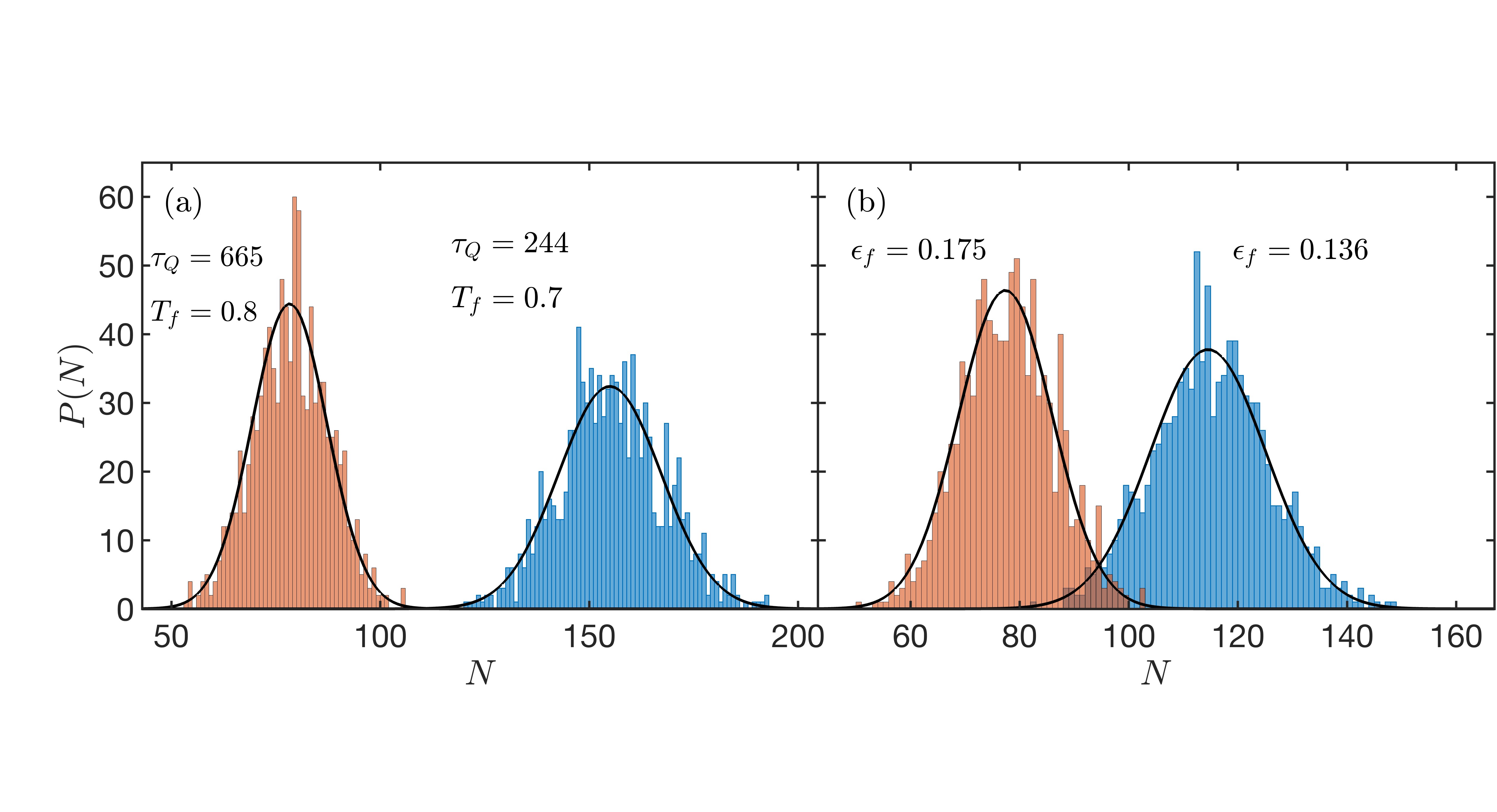}
    \caption{Vortex loop number histograms. 
    (a) Slow driving limit for $(T_f=0.7T_c,\tau_Q=244)$ and $(T_f=0.7T_c,\tau_Q=665)$. (b) Fast quench regime for $\tau_Q=1$ and for $\epsilon_f=0.096$ and $\epsilon_f=0.158$. Both regimes are well described by the black Gaussian distributions, with mean and variance determined numerically.}
    \label{fig4}
\end{figure}

In the limit of large $\mathcal{N}$, the binomial distribution converges to a Gaussian by the central-limit theorem. Once the mean and variance are set by the scaling laws discussed above, the full distribution is therefore expected to exhibit a universal collapse. This is borne out by our simulations. For slow quenches, the vortex-loop histograms in Fig.~\ref{fig4}(a) are well captured by Gaussian fits using the measured values of $\kappa_{1,2}(N)$ at $T_f = 0.7T_c$ and $0.8T_c$. The same holds in the fast-quench regime: as shown in Fig.~\ref{fig4}(b), the histograms at $\tau_Q = 1$ for quench depths $\epsilon_f = 0.136$ and $0.175$ closely follow Gaussian distributions with the numerically observed mean and variance. 

Together, the scaling of cumulants and the emergence of Gaussian statistics provide strong evidence that vortex formation in this system obeys universal behavior characteristic of extended defects in three dimensions, across both slow and rapid quenches. Further discussion of the predicted universal freeze-out scaling in both regimes is provided in ~\cite{Supplementary}.

Beyond the statistics of vortex number, the total vortex-loop length $L$ provides an independent observable that captures correlations intrinsic to extended defects, absent in systems dominated by point-like excitations. The binomial picture for the loop number can be refined by invoking the Gaussian random-walk statistics of closed string loops~\cite{Tanmay1984,Antunes1998,Copeland1998,Nemirovskii2006}, which govern the distribution of loop lengths. Vortex loops form via independent random walks, with lengths $\ell = n \ell_0$ set by first-return statistics. Here, $q$ denotes the step probability, and the effective lattice spacing $\ell_0 \propto \hat{\xi}$ encodes the nonequilibrium correlation scale. The resulting loop-length distribution follows
\begin{eqnarray}
P(\ell) \propto q^{1/2} (\ell_0 n)^{-5/2} \propto \ell^{-5/2},
\end{eqnarray}
valid for $1\ll n \ll L^2$~\cite{Austin_PRD1994}. The formation probability $p$ is determined by the closure condition of the walk, giving $p \equiv \sum_{n=1}^{\infty} P(\ell = n\ell_0) \sim\sqrt{q}$ for $q \ll 1$; as shown in~\cite{Supplementary}.
\begin{figure}[t]
    \centering    
        \includegraphics[width=0.99\linewidth,trim={0 4.cm 0 350 cm},clip]{ 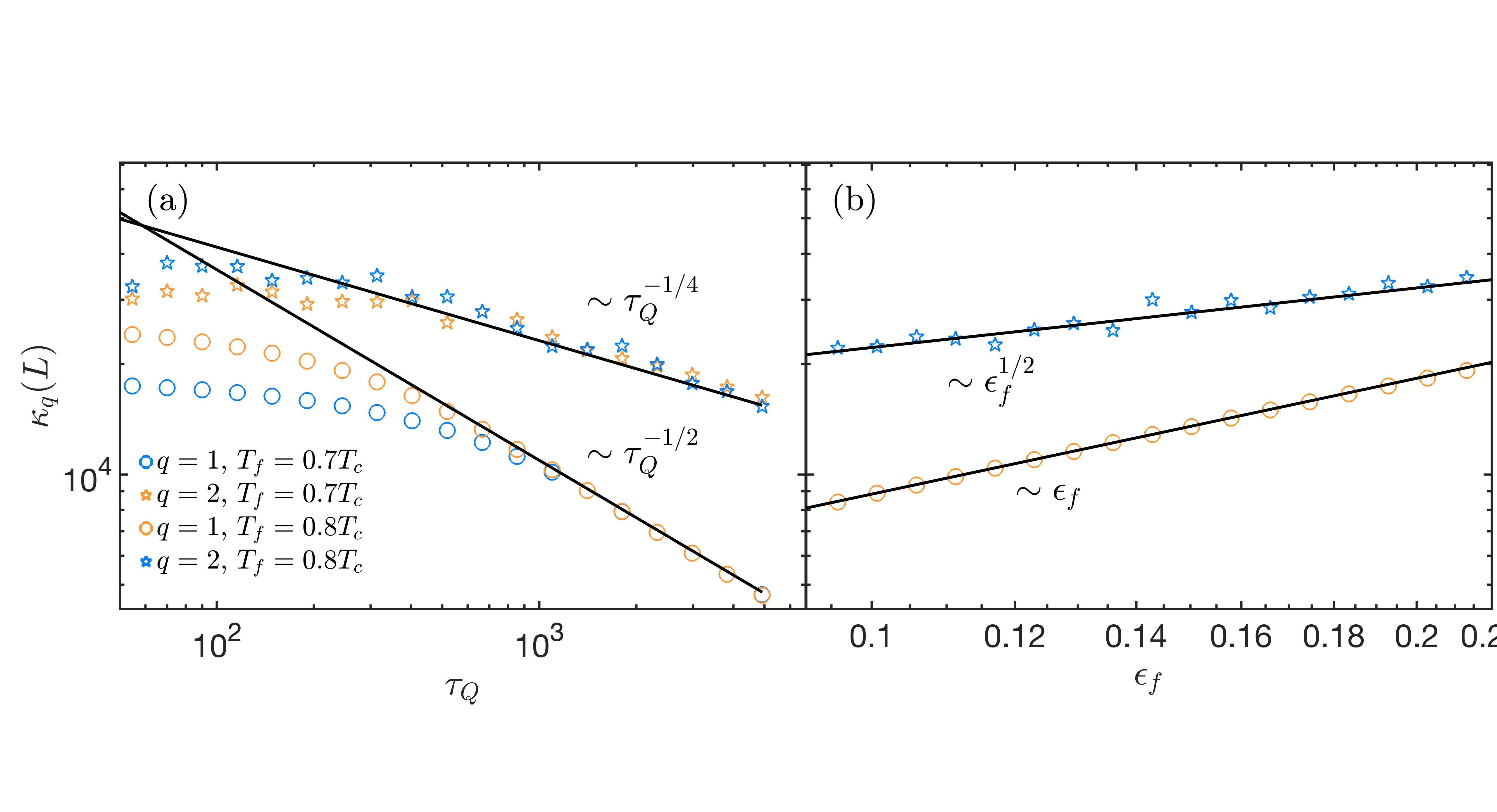}
    \caption{
        (a) Scaling of the first two vortex loop length cumulant, exhibiting approximately the same power-law behavior for the two final temperatures, $T_f=0.7T_c,\,0.8T_c$,
        $\kappa_{1}=(39700\pm900) \tau_Q^{-0.522\pm0.003},\quad \kappa_{2}=(13500\pm440) \tau_Q^{-0.255\pm0.004}$. (b) Similar good matching with the theoretical predictions in fast quenches with the best fit giving  $\kappa_{1}(L)=(99200\pm1000)\epsilon_f^{1.051\pm0.006},\quad\kappa_{2}(L)=(77000\pm900)\epsilon_f^{0.54\pm0.007}$.
    }
    \label{fig:L_cumulants}
\end{figure}

The cumulants of the total vortex-loop length $L$ exhibit universal power-law scaling that extends beyond simple proportionality. As vortex loops contribute independently, the total length cumulants are given by the product of the average loop number and the single-loop length cumulants, $\kappa_q(L)\propto \kappa_1(N)\kappa_q(\ell)$ \cite{Supplementary}. This factorization accurately captures the first two cumulants, with subleading loop–loop correlations leaving the scaling exponents unchanged:
\begin{eqnarray}
&&\kappa_1(L)\propto \tau^{-1/2}_Q,\quad\kappa_2(L)\propto \tau^{-1/4}_Q,\\
&&\kappa_1(L)\propto \epsilon_f,\,\,\qquad\kappa_2(L)\propto \epsilon^{1/2}_f.
\end{eqnarray}
This is confirmed in Fig.~\ref{fig:L_cumulants}(a,b) for slow and fast quenches, respectively.

\begin{figure}
    \centering
    \includegraphics[width=0.99\linewidth,trim={0 6cm 0 12cm},clip]{ 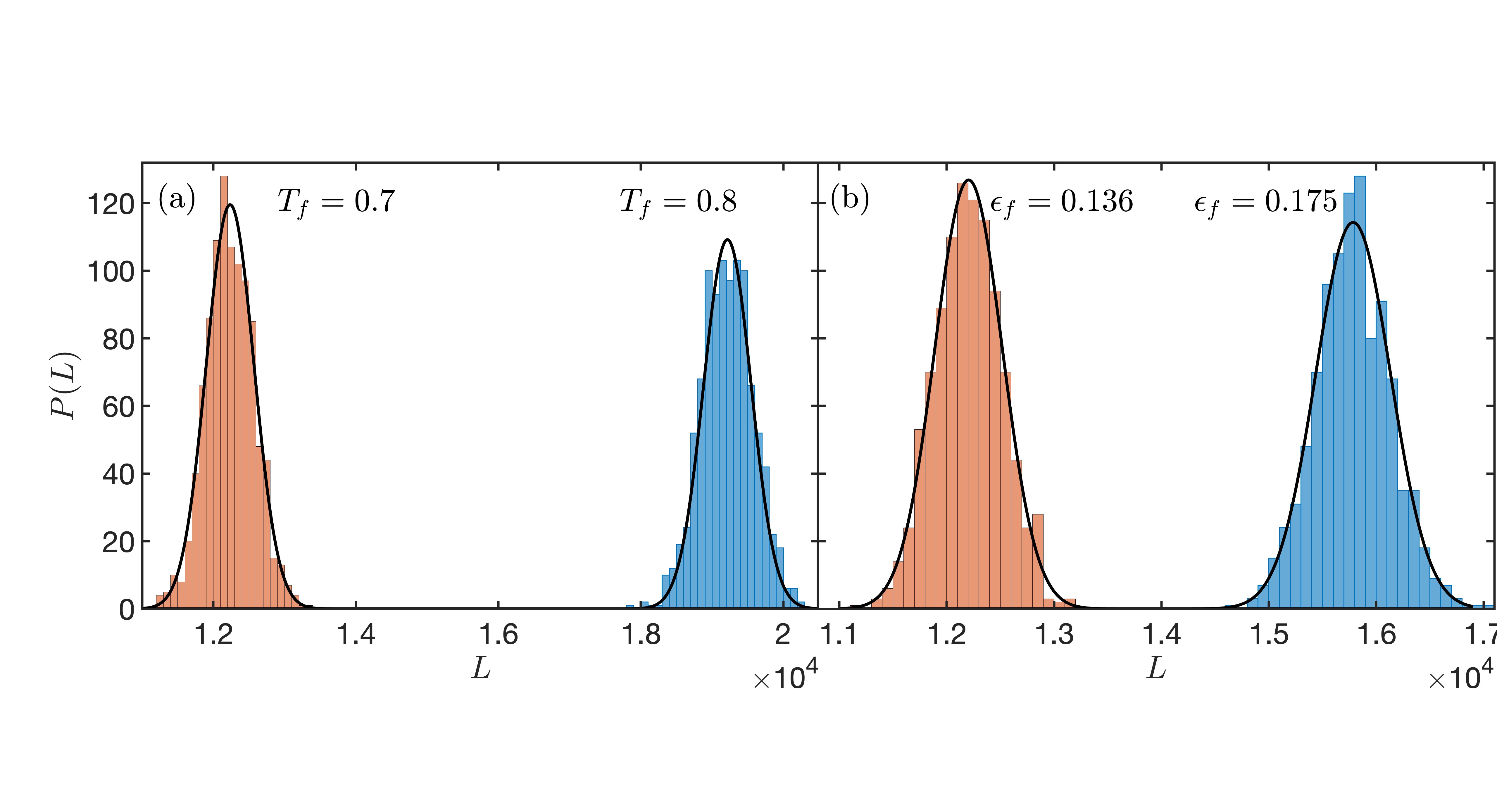}    
    \caption{Vortex loop length histograms. 
    (a) Slow driving limit for $(T_f=0.7T_c,\tau_Q=244)$ and $(T_f=0.7T_c,\tau_Q=665)$, showing good matching with a Gaussian distribution with the numerically obtained parameters. (b) Fast quench regime for $\tau_Q=1$ and for $\epsilon_f=0.096$ and $\epsilon_f=0.158$.}
    \label{fig:L_totStat}
\end{figure}

Despite the nontrivial length distribution $P(\ell)$ and non-proportional single-loop cumulants, the total vortex-line statistics are governed by the central limit theorem. Due to the independence of individual loops, the distribution rapidly converges to a Gaussian, as shown in Fig.~\ref{fig:L_totStat} for slow and fast quenches, using the same parameters as for the vortex loop numbers.

\begin{figure}[h]
    \centering
    \includegraphics[width=0.99\linewidth]{ 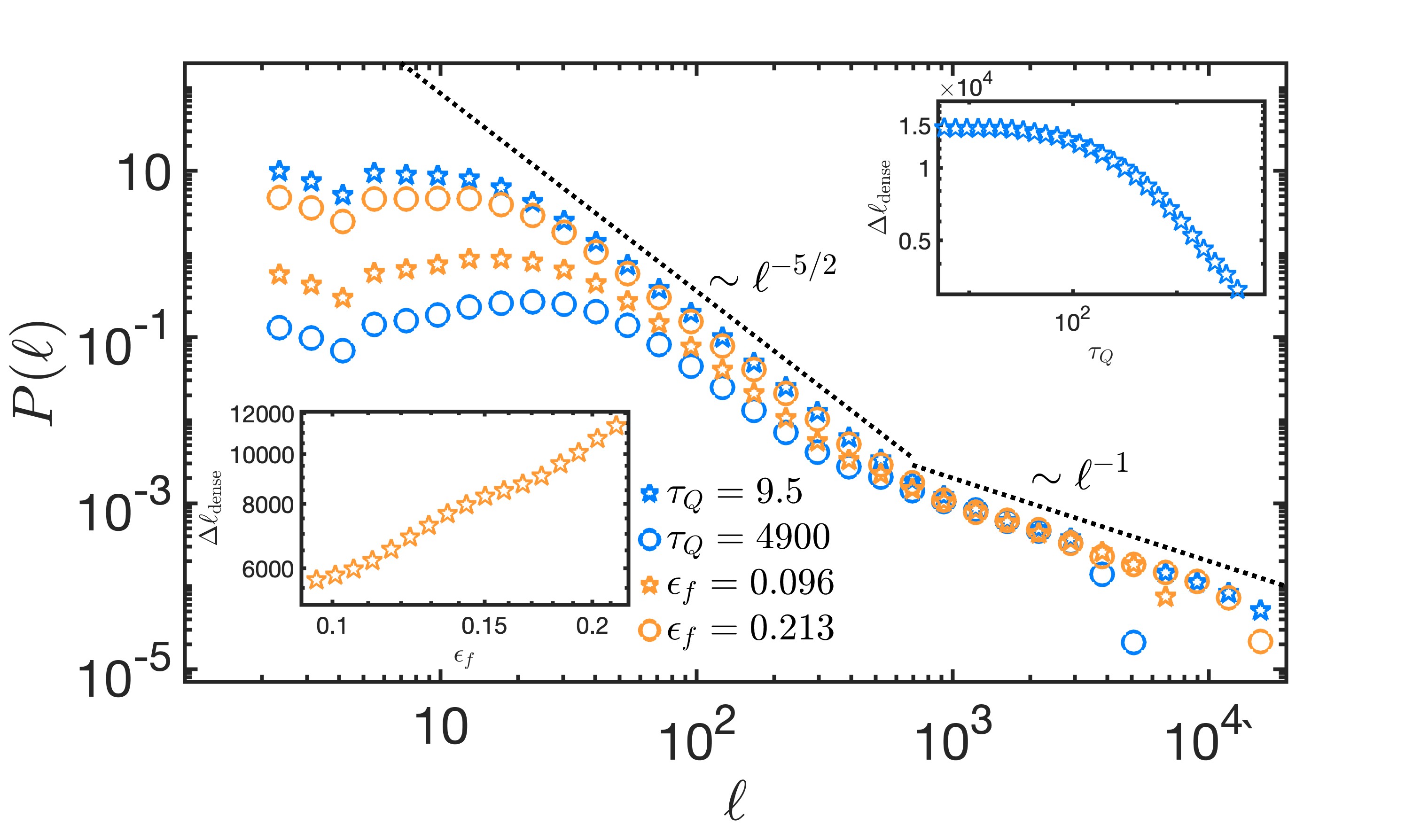}
    \caption{Single loop length distribution exhibiting an initial plateau followed by a universal power-law scaling
       captured by $P(\ell)\propto \ell^{-2.41\pm0.005}$ consistent with the universal $3D$ random walk description.
     A crossover takes place to the densely packed limit with a slower decay $P(\ell)\propto \ell^{-1.01\pm0.03}$ for long loops, $\ell\sim L^2$. The top inset shows the shrinking of the $1/\ell$ scaling regime with increasing $\tau_Q$, while the bottom inset shows the fast quench counterpart.
    }
    \label{fig:P_l}
\end{figure}

Finally, we show the single-loop length distributions in Fig.~\ref{fig:P_l} for both slow ($\tau_Q=9.5,4900$) and fast quenches ($\epsilon_f=0.096,0.213$). For sufficiently long loops, $\ell \gtrsim \ell_0^3 q$, the distribution follows the universal power law $P(\ell)\sim \ell^{-5/2}$, in perfect agreement with previous studies of vortex string dynamics, where a kinetic master equation exploiting a similar Gaussian random-walk description of the loops was used~\cite{Nemirovskii2005,Nemirovskii2006,Nemirovski_PRB_2008}. For shorter loops, an initial plateau appears, reflecting the breakdown of the leading-order first-return approximation;  see~\cite{Supplementary}.

The $3D$ random-walk picture also accurately accounts for a different $1/\ell$ scaling behavior in the densely packed limit where $n\gg L^2$, as shown in~\cite{Supplementary}. This behavior was observed in $U(1)$ gauge theories~\cite{Kobayashi2016,Kobayashi16b}, densely packed percolation models~\cite{Melchert2011,Tsubota_1_l_2006}, and derived analytically using a statistical physics approach in Ref.~\cite{Austin_PRD1994}.
The crossover length scale can be estimated by the condition of the loops filling up the available volume, yielding $\ell_{\mathrm{cross}}\sim L^2$ within the $3D$ random walk picture, while they cannot grow longer than $\ell_\mathrm{max}\sim L^3$.
As a result, the probability weight of this $1/\ell$ scaling regime becomes subleading compared to $1/\ell^{5/2}$ as $\Delta\ell_{\mathrm{dense}}=\ell_\mathrm{max}-\ell_\mathrm{cross}\sim\log L/L^3$; see~\cite{Supplementary}.
Finally, we analyze how the slow and fast dynamics affect the statistics of the vortex tangle. For small $\tau_Q$-s and large enough $\epsilon_f$-s, the independence of the vortex line formations is restricted by the large number of individual loops,  leading to a slight extension of the $1/\ell$ scaling regime. These features are shown in the inset of Fig.~\ref{fig:P_l}, reflecting the diminishing degree of the relative extension of the $1/\ell$ regime, $\Delta\ell_\mathrm{dense}$, with increasing $\tau_Q$ and decreasing $\epsilon_f$.

In summary, we have investigated the dynamics of vortex string formation after a temperature quench in a 3$D$ holographic superfluid. Beyond the universal KZM scaling of loop numbers, we have uncovered statistical signatures of extended defects: loop lengths follow the distribution $P(\ell)\sim \ell^{-5/2}$ from 3$D$ random walk return statistics, and vortex length cumulants display distinct universal power-law scaling. These results highlight a new regime of defect dynamics unique to extended objects, with implications for dense-tangle crossovers and broader applications in driven superfluids, turbulence, and strongly coupled systems.
As our predictions apply to generic 3$D$ $U(1)$ symmetry-breaking scenarios and are not restricted to the holographic setting, they can be tested experimentally, e.g., using 3$D$ superfluid Bose and Fermi gases \cite{Ko2019,Goo2021,lee2024universal} and multiferroics \cite{lin2014topological,2023Du,Najeeb2025}.

{\it Acknowledgments.---} 
H. B. Z. acknowledges the support from the National Natural Science Foundation of China (under Grant No. 12275233 and 12547103).  AG and AdC acknowledge financial support from the Luxembourg National Research Fund under Grants No. C22/MS/17132060/BeyondKZM and C22/MS/17132054/AQCQNET.


\bibliography{main}
\clearpage
\onecolumngrid
\begin{appendix}
\begin{center}
\renewcommand\thefigure{S\arabic{figure}}    
\setcounter{figure}{0} 
\renewcommand{\theequation}{S\arabic{equation}}
\setcounter{equation}{0}
\renewcommand{\thesubsection}{SI\arabic{subsection}}
\end{center}
\setcounter{equation}{0}
\setcounter{table}{0}


\section{AdS$_5$ black hole, bulk action and detail of holographic mapping}
\label{appendixA}

In this paper, we use the standard holographic superfluid model \cite{Gubser2008,Hartnoll2008}, defined in the background of an AdS$_5$ black hole. The bulk theory is the Abelian-Higgs-Einstein model with a local $U(1)$ gauge field $\textbf{A}$ and a charged scalar field $\Psi$ coupled to the AdS black hole, with the action 
\begin{equation}
S=\int d^5x \sqrt{-g}\Big[R+ \frac{\Lambda}{\ell^2}+\frac{1}{q^2}\big(-\frac{1}{4}F^2-( |D\Psi|^2-m^2|\Psi|^2)\big)\Big].
\end{equation}
$R$ and $\Lambda$ are the Ricci scalar and the negative cosmological constant, respectively. According to the standard AdS/CFT dictionary,
the conserved boundary current $J_M(x,u)$ is mapped to the dynamical $U(1)$
gauge field $A_M(x, u)$ in the gravitational bulk, while the scalar operator $\psi$ is mapped to
a bulk scalar field $ \Psi$.
In units $\hbar = c = G_N = 1$, when the Abelian-Higgs model with only a quadratic potential of the scalar field is defined in a
flat space-time, and there is no symmetry broken, a quartic potential is needed. However, when the charged scalar field is coupled to gravity with a negative cosmological constant, the scalar field will condense, as a stable finite value solution, when the black hole
temperature is below a critical value \cite{Gubser2008}.

Following \cite{Hartnoll2008}, we work in the probe limit, which applies when the charge $q$ of $\Psi$ is
large. In this limit, the back-reaction from the matter fields can be ignored.
The gravitational system is then approximated by an Abelian-Higgs model
defined in a Schwarzschild black hole background geometry
\begin{equation}
ds^2=\frac{\ell^2}{u^2}(-f(u) dt^2 -2Dtdu+ dx^2+dy^2+dz^2),
\end{equation}
where $f(u)=1-(u/u_h)^4$, $u$ is the extra bulk dimension,
$u_h$ is the horizon, while $u=0$ is the AdS boundary.
The black hole's Hawking temperature $T$ is proportional to
$u_h$, and there is a critical $T_c$ below which the scalar field
will condense. Without loss of generality, we set $\ell=1$.

The equations of motion of the fields $A_M$ and $\Psi$ read
\begin{equation}
d_{N} F^{M,N}=J^{N},\quad (-D^2+m^2) \Psi=0,
\end{equation}
in the axial gauge $A_{u}=0$, fixing the gauge degree of freedom. This yields highly non-linear
coupled PDEs for the five fields $A_t, A_x, A_y, A_z, \Psi$.
The equations of motion can be written
explicitly as
\begin{align}
\label{eomf1}
m^2\Psi+3u(i A_t\Psi+f\partial_u\Psi-\partial_t \Psi)+u^2 [\Psi (A_x^2+A_y^2+A_z^2+i(-\partial_u A_t+\partial_x A_x+\partial_y A_y+\partial_z A_z))+\nonumber\\
2i(- 
A_t\partial_u\Psi+A_x\partial_x\Psi+A_y\partial_y\Psi+A_z\partial_z\Psi)-\partial_x^2\Psi -\partial_y^2\Psi-\partial_z^2\Psi-\partial_u f \partial_u \Psi-f \partial_u^2 \Psi+2 \partial_t \partial_u \Psi]=0,
\end{align}
\begin{align}
\label{eomf2}
u \partial_u A_t+u^2 \partial_u(-\partial_u A_t+ \partial_x A_x+ \partial_y A_y+ \partial_z A_z)+i \Psi^* \partial_u \Psi-i \Psi \partial_u \Psi^*=0,
\end{align}
\begin{align}
\label{eomf3}
 2 A_x |\Psi|^2+i \Psi^* \partial_x \Psi-i \Psi \partial_x \Psi^*+u (\partial_x A_t-\partial_t A_x+f\partial_u A_x)-u^2[\partial_u(f\partial_u A_x) +\partial_y^2 A_x+\partial_z^2 A_x-\partial_x( \partial_y A_y+ \partial_z A_z)\nonumber\\
+\partial_u \partial_x A_t-2 \partial_t \partial_u A_x ]=0,
\end{align}
\begin{align}
\label{eomf4}
 2 A_y |\Psi|^2+i \Psi^* \partial_y \Psi-i \Psi \partial_y \Psi^*+u (\partial_y A_t-\partial_t A_y+f\partial_u A_y)-u^2[\partial_u(f\partial_u A_y) +\partial_z^2 A_y+\partial_x^2 A_y-\partial_y( \partial_z A_z+ \partial_x A_x)\nonumber\\
+\partial_u \partial_y A_t-2 \partial_t \partial_u A_y ]=0,   
\end{align}
\begin{align}
\label{eomf5}
 2 A_z |\Psi|^2+i \Psi^* \partial_z \Psi-i \Psi \partial_z \Psi^*+u (\partial_z A_t-\partial_t A_z+f\partial_u A_z)-u^2[\partial_u(f\partial_u A_z) +\partial_x^2 A_z+\partial_y^2 A_z-\partial_z( \partial_x A_x+ \partial_y A_y)\nonumber\\
+\partial_u \partial_z A_t-2 \partial_t \partial_u A_z ]=0,   
\end{align}
\begin{align}
\label{eomf6}
2 A_t |\Psi|^2 + i \Psi^* \partial_t \Psi-i \Psi \partial_t \Psi^*
+f (-i \Psi^* \partial_u \Psi+i \Psi \partial_u \Psi^*) - u^2 [\partial_x^2 A_t+\partial_y^2 A_t+\partial_z^2 A_t+f\partial_u(\partial_x A_x+\partial_y A_y+\partial_z A_z)\nonumber \\
-\partial_t(\partial_u A_t+\partial_x A_x+\partial_y A_y+\partial_z A_z)]=0.
\end{align}
The six partial differential equations \eqref{eomf1}-\eqref{eomf6} are not independent, and it suffices to choose five of them. In this work, we choose Eqs. 
\eqref{eomf1}-\eqref{eomf5}, using the remaining Eq. \eqref{eomf6} to check self-consistency.

At the horizon, in our ingoing coordinates, physical solutions should be
regular. Near the boundary, a general solution takes the following form
\begin{equation}
A_\nu(t,x,y,z,u)=a_{\nu}(t,x,y,z)+ O(u),  \quad \Psi(t,x,y,z,u)= \Psi^- u^{\Delta^-} +  \Psi^+ u^{\Delta^+}.
\end{equation}
where
\begin{equation}
\Delta^\pm = \frac{4 \pm \sqrt{16 + 4 m^2}}{2}.
\end{equation}
We take $m^2=-3$, and note that 
$a_\nu$ defines a background gauge field for the $U(1)$ current $j^{\nu}$ of the dual theory, with $\Psi^-$
an external source for the condensate $\Psi^+ = \la \mathcal{O} \ra$, where $\psi$ is the operator dual to the scalar
field.
Due to the  scaling symmetry of the equations of motion, the  temperature is proportional $1/\mu$, which means that we can
set $u_h=1$, and effectively reduce the temperature to induce a superfluid phase transition by increasing $\mu$.

We are interested in the spontaneously broken superfluid phase with finite
chemical potential, and the external sources $\Psi^-$ must be set to zero at the boundary,
\begin{equation}
a_t(t,x,y,z)=\mu, \quad \textbf{a}=0,  \quad  \Psi^-=0.
\end{equation}
The expectation value of the superfluid condensation is determined by the subleading asymptotics of $\Psi$
\begin{equation}
\langle \mathcal{O} \rangle=\frac{1}{6}\partial_u^3 \Psi|_{u=0}.
\end{equation}
This theory has a critical chemical potential value $\mu_c=4.16$. The normal and superfluid states correspond to chemical potentials below and above this value, respectively.


\section{Method of quenching the three-dimensional superfluid and numerical details of solving PEDs }
\label{appendixB}

From dimensional analysis, we know that the temperature of the black hole $T$ has mass dimension one, the same as the mass dimension of the chemical potential $\mu$, so $T/\mu$ is dimensionless. From the scaling symmetry of the EOMs, it is easy to find that decreasing the temperature is equivalent to increasing the chemical potential. 
Therefore, we can actually quench the chemical potential 
\begin{eqnarray}
\mu(t)=
\begin{cases}
\mu_c(1-t/\tau_Q)^{-1}  & 0 \leq t<t_f \\
\mu_f              & t \geq t_f
\end{cases}
\end{eqnarray} 
to achieve the linear cooling process 
\begin{eqnarray}
T(t)=
\begin{cases}
T_c(1-t/\tau_Q)  & ~~~0 \leq t<t_f \\
T_f              & ~~~t \geq t_f
\end{cases},
\end{eqnarray} 
where $\tau_Q$ is the quench time of a linear quench. The quench end time $t_f$ is defined from $T(t_f)=T_f$
and is given by $t_f=\tau_Q(1-T_f/T_c)$.
Before the linear thermal quench, we prepare the normal fluid at $T_c$. Numerically, we start with a set of normal state solutions that are uniform in the $(x,y,z)$ directions and written as
\begin{equation}
 A_t =\mu_c(1-u^2),~~~A_x=A_y=A_z=\Psi=0.
\end{equation}
Then, we add random noise to $\Psi$  to simulate the thermal fluctuation of the system, that is, $S(x,y,z,u)$.
In this way, the normal fluid that grows at the boundary also inherits random noise $s(x,y,z)$. Noise $s(x,y,z)$ satisfies the normal distribution, with a vanishing statistical average $\langle s(x,y,z) \rangle=0$ and two-point correlations $\langle s(x,y,z)s(x', y', z') \rangle= h  \delta(x-x',y-y',z-z')$, where the amplitude $h={10}^{-4}$. After a little heating, the fluid disk is ready for quenching.  
We count the number and length of the vortex strings at the freeze-out time $\hat{t}$.
Numerically,  pseudo-spectral methods and the fourth-order Runge-Kutta method are adapted to solve the highly nonlinear bulk equations of motion (PDEs).  Specifically, $21$ Chebyshev grid points are used in the $u$-direction, and $151$ Fourier grid points are used in each of the $(x, y, z)$ spatial directions to discretize the entire bulk.  Then, we select $0.05$ as the single time step size and apply the boundary conditions in the $u$ direction at each step of the time evolution, thus completing the numerical simulation of the holographic superfluid system.

\section{3$D$ random walk framework for the vortex line length statistics}

In this section, we discuss the first-return probability of a uniform 3$D$ random walk and show how to compute the first two length cumulants. 
Denoting by $P_n(\textbf r)\equiv \mathrm{Prob}(\textbf X_n=\textbf r)$ the probability for the random walk being at $\textbf r=(x,y,z)$ after the $n$-th step, the probability of one step is given by 
\begin{eqnarray}
&&P_{n+1}(\textbf r)=(1-q)P_n(\textbf r)\\
&&+\frac{q}{6}\left[P_{n}((x+\ell_0,y,z))+P_{n}((x-\ell_0,y,z))+P_{n}((x,y+\ell_0,z))+P_{n}((x,y-\ell_0,z))+P_{n}((x,y,z+\ell_0))+P_{n}((x,y,z-\ell_0))\right].\nonumber
\end{eqnarray}
First, we express the characteristic function of a single step as 
\begin{eqnarray}
    \phi(\mathbf k)=\mathbb E[e^{i\mathbf k\mathbf r_1}]=(1-q)+\frac{q}{3}\left[\cos(k_x\ell_0)+\cos(k_y\ell_0)+\cos(k_z\ell_0)\right].
\end{eqnarray}
When the random walk starts from the origin, the probability of arriving at an arbitrary position $\textbf r$ after $n$ steps can be expressed as
\begin{eqnarray}
    P_n(\textbf r)=\int_{-\pi}^\pi\frac{\mathrm d^3\mathbf k}{(2\pi)^3}\phi^n(\textbf k) e^{-i\textbf k\textbf r}\rightarrow P_n(\textbf0)=
    \int_{-\pi}^\pi\frac{\mathrm d^3\mathbf k}{(2\pi)^3}\phi^n (\textbf k),
\end{eqnarray}
where in the second expression $\textbf r=0$ has been set to express the return probability. However, the first return probability $F_n$ is generally harder to capture as it can only be expressed by a self-consistent equation with $P_n(\textbf 0)\equiv P_n$,
\begin{eqnarray}
    P_n=\sum_{m=1}^n F_mP_{n-m},\quad\tilde P(z)=\frac{1}{1-\tilde F(z)},\quad\tilde P=\sum_{n=0}^\infty P_nz^n,\quad\tilde F=\sum_{n=1}^\infty F_nz^n,\quad\lvert z\rvert<1,
\end{eqnarray}
    where the last expression is formulated in terms of the generating functions for $G_n=P_n,\,F_n$. From here, the first return probability is expressed via its generating function as
    \begin{eqnarray}
            \tilde P(z)=\sum_{n=0}^\infty z^n \int_{-\pi}^\pi\frac{\mathrm d^3\textbf k}{(2\pi)^3} \phi^n(\textbf k)=\int_{-\pi}^\pi\frac{\mathrm d^3\textbf k}{(2\pi)^3}\frac{1}{1-z\phi(\textbf k)},\quad\tilde F(z)=1-\left[\int_{-\pi}^\pi\frac{\mathrm d^3\textbf k}{(2\pi)^3}\frac{1}{1-z\phi(\textbf k)}\right]^{-1}.
    \end{eqnarray}
    The first return probability can thus be expressed by the Cauchy theorem as
    \begin{eqnarray}
        F_n=\frac{1}{2\pi i}\oint_{\lvert z\rvert=\epsilon>0}\mathrm dz\frac{\tilde F(z)}{z^{n+1}}=\frac{1}{2\pi i}\oint_{\lvert z\rvert=\epsilon}\mathrm dz\frac{1-\left[ \int_{-\pi}^\pi\frac{\mathrm d^3\textbf k}{(2\pi)^3}\frac{1}{1-z\phi(\textbf k)}\right]^{-1}}{z^{n+1}},
    \end{eqnarray}
    where the complex contour integration is performed along a circle of radius $\epsilon>0$ containing the origin $z=0$.
    The large $n$ expansion in the limit $n\ll L^2$ is identical to the small $k$ expansion, $\phi(\textbf k)\approx 1-\frac{q\ell^2_0}{6}\textbf k^2$, by which the momentum integral becomes
    \begin{eqnarray}\label{eq: P_z_expansion}
        &&\int\frac{\mathrm d^3\textbf k}{(2\pi)^3}\frac{1}{(1-z)+\frac{q\ell^2_0}{6}z\textbf k^2}\approx \int_{-\pi}^\pi\frac{\mathrm d^3\textbf k}{(2\pi)^3}\frac{1}{1-\phi(\textbf k)}-\frac{1}{(2\pi)^3}\int\mathrm d^3\mathbf k\frac{(1-z)}{\frac{q\ell^2_0}{6}z\mathbf k^2\left[(1-z)+\frac{q\ell^2_0}{6}z\mathbf k^2\right]}\nonumber\\
        &&\approx\tilde P(1)-\frac{1}{4\pi}\frac{(1-z)^{1/2}}{\left(zq\ell^2_0/6\right)^{3/2}},\,\tilde P(1)\approx \int\frac{\mathrm d^3\textbf k}{(2\pi)^3}\frac{1}{q\left[1-\left(\cos(k_x\ell_0)+\cos(k_y\ell_0)+\cos(k_z\ell_0)\right)/3\right]}=1.516\,q^{-1},
    \end{eqnarray}
    which is justified from the upcoming complex contour integrals, showing that all $(1-z)$ terms contribute with subleading factors of $1/n$. This implies that the correct leading order expansion is indeed up to $\mathbf k^2$ in the limit $n\lesssim L^2$. Here, we used the full expression for $\tilde P(1)$ rather than the leading order expansion.
    For large $n$, only the singular term matters in $\tilde F(z)$,
    \begin{eqnarray}
        \tilde F(z)\approx1-\frac{1}{\tilde P(1)-\frac{1}{4\pi}\frac{(1-z)^{1/2}}{\left(zq\ell^2_0/6\right)^{3/2}}}\approx 1-\frac{1}{\tilde P(1)}-\frac{1}{\tilde P^2(1)}\frac{1}{4\pi}\frac{(1-z)^{1/2}}{\left(zq\ell^2_0/6\right)^{3/2}},
    \end{eqnarray}
    from where the Cauchy integral becomes
    \begin{eqnarray}
        F_n\approx\frac{1}{2\pi i}\oint_{\lvert z\rvert=\epsilon}\frac{ 1-\frac{1}{\tilde P(1)}-\frac{1}{\tilde P^2(1)}\frac{1}{4\pi}\frac{(1-z)^{1/2}}{\left(zq\ell^2_0/6\right)^{3/2}}}{z^{n+1}}\approx\frac{1}{2}\left(\frac{3}{2\pi q}\right)n^{-3/2}=\frac{1}{8\pi^{3/2}\tilde P^2(1)}\left(\frac{q\ell^2_0}{6}\right)^{-3/2}n^{-3/2}\propto q^{1/2} n^{-3/2}.\qquad\,
    \end{eqnarray}

     Notice that the leading order $1/n$ power originates from the $\sqrt{1-z}$ term. Thus, for every higher order of $(1-z)$ entering the expression would only provide subleading higher orders, $n^{-5/2},\,n^{-7/2},\dots$. This justifies the leading order expansion in Eq.~\eqref{eq: P_z_expansion} in the limit of $n\lesssim L^2$.
    In the realistic vortex formation scenario, one also has to account for the probability of choosing a given starting point for the random walk, i.e., divide by an additional factor of $n$, giving
    \begin{eqnarray}
        P(\ell=n\ell_0)=L^3\frac{F_n}{n}\sim L^3\ell^{-5/2}.
    \end{eqnarray}
    Furthermore, we verify that the next-to-leading order correction by the quartic expansion of the characteristic function is indeed governed by the extra power of $1-z$ and a corresponding subleading $1/n$ factor,
    \begin{eqnarray}
        \phi(\textbf k)&&\approx 1-\frac{q\ell^2_0}{6}\textbf k^2+\frac{q\ell^4_0}{72}(k^4_x+k^4_y+k^4_z)\Rightarrow \int\frac{\mathrm d^3\textbf k}{(2\pi)^3}\frac{1}{1-z\phi(\textbf k)}\approx \int\frac{\mathrm d^3\textbf k}{(2\pi)^3}\frac{1}{(1-z)+z\frac{q\ell^2_0}{6}\textbf k^2+\frac{q\ell^4_0}{72}(k^4_x+k^4_y+k^4_z)}\\
        &&\approx -\frac{1}{\tilde P^2(1)}\frac{1}{4\pi}\frac{(1-z)^{1/2}}{\left(\frac{zq\ell^2_0}{6}\right)^{3/2}}\left[1-\frac{\ell^2_0}{12}(1-z)\right]+\mathrm{const.},\nonumber
    \end{eqnarray}
    where the additional constant term does not contribute to the final leading order result.
    The corresponding Cauchy integral becomes
    \begin{eqnarray}
        F_n&&=\frac{1}{2\pi i}\oint_{\lvert z\rvert=\epsilon}-\frac{\frac{1}{\tilde P^2(1)4\pi}(1-z)^{-1/2}\left(\frac{zq\ell^2_0}{6}\right)^{-3/2}\left[1+\frac{\ell^2_0}{12}(1-z)\right]}{z^{n+1}}\nonumber\\
        &&=\frac{1}{8\pi^{3/2}\tilde P^2(1)}\left(\frac{q\ell^2_0}{6}\right)^{-3/2}n^{-3/2}\left[1-\frac{\ell^2_0}{12}\frac{1}{n}\right]
        \nonumber\\
        &&\propto q^{1/2}\,n^{-3/2}\left(1-\frac{\ell^2_0}{12}\frac{1}{n}\right),
    \end{eqnarray}
    showing that the correction becomes negligible for $n\gg \ell^2_0\propto \hat\xi^2$, as stated in the main text. This implies that the larger the freeze-out length scale, the fewer loops are formed on average, and consequently, a larger number of line elements are required for the leading-order continuum approximation to hold.

    Next, we provide the analytical derivation of the $1/\ell$ scaling regime based on the above framework of $3D$ random walks. This regime arises because loops reach the length scale of the cube. As a result, the integral approximations over the momenta of the characteristic function $\phi(\mathbf k)$ no longer hold. In particular, one needs to restore the original discrete values, $\mathbf k=2\pi/L(n_x,n_y,n_z),\,n_{x,y,z}=-2L,-2L+1\dots2L-,2L$ and use that $1-z\sim 1/n\ll L^{-2}$
     \begin{eqnarray}
         \sum_{\mathbf k}\frac{1}{1-z\phi(\mathbf k)}=\sum_{\mathbf k}\frac{1}{1-z-z(\phi(\mathbf k)-1)}=\frac{1}{1-z}\sum_{\mathbf k}\frac{1}{1+\frac{z(1-\phi(\mathbf k))}{1-z}}\approx\frac{1}{z}\sum_{\mathbf k\neq 0}\frac{1}{1-\phi(\mathbf k)}+\frac{1}{1-z}.
     \end{eqnarray}
     Thus, the first return probability is given by
     \begin{eqnarray}
         F_n=\frac{1}{2\pi i}\oint_{|z|=\epsilon}\mathrm dz\,\frac{1-\frac{1}{\frac{1}{L^3z}\sum_{\mathbf k\neq 0}\frac{1}{\phi(\mathbf k)-1}+\frac{1}{L^3(1-z)}}}{z^{n+1}}=\frac{L^3}{2\pi i}\oint_{|z|=\epsilon}\frac{1}{\frac{1}{z}C-\frac{1}{1-z}}\frac{1}{z^{n+1}}=\frac{L^3}{(1+C)^2}\left(1+\frac{1}{C}\right)^{n-1}\sim \frac{1}{L^3},
     \end{eqnarray}
     where $C\equiv\sum_{\mathbf k}\frac{1}{\phi(\mathbf k)-1}\approx 1.516L^3q^{-1}$ and we used that $L^3\gg n$ as constrained by the dimensionality of the system. By this, the last term was dropped as $\left(1-\frac{1}{C}\right)^{n-2}\approx e^{-n/C}\sim e^{n/L^3}\approx 1$.

     Together with the overall arbitrariness of the starting position accounted for by the factor of $L^3/n$, we obtain the following
     \begin{eqnarray}
         P(\ell)\sim 1/\ell.
     \end{eqnarray}

     Notice that going further with one step in  
     \begin{eqnarray}
     \sum_{\mathbf k}\frac{1}{1-z\phi(\mathbf k)}&=&\sum_{\mathbf k}\frac{1}{1-z-z(\phi(\mathbf k)-1)}=\frac{1}{1-z}\sum_{\mathbf k}\frac{1}{1+\frac{z(1-\phi(\mathbf k))}{1-z}}\approx\frac{1}{z}\sum_{\mathbf k\neq 0}\frac{1}{1-\phi(\mathbf k)}-\frac{1-z}{z^2}\sum_{\mathbf k\neq 0}\frac{1}{(1-\phi(\mathbf k))^2}+\frac{1}{1-z}\nonumber\\
     &=&-\frac{1}{z}C-\frac{1-z}{z^2}C_2+\frac{1}{1-z},
     \end{eqnarray}
     with $C\sim L^3,\,C_2\sim L^4$, results in a subleading correction in $F_n$. In particular,
     \begin{eqnarray}
         F_n&=&\frac{L^3}{2\pi i}\oint_{|z|=\epsilon}\frac{1}{\frac{1}{z}C-\frac{1}{1-z}+\frac{1-z}{z^2}C_2}\frac{1}{z^{n+1}}\\
         &=&\frac{2^{-n-1}}{(1+C-C_2)^2\sqrt{C^2+4C_2}}
\Bigg[
\left(\frac{-C+2C_2+\sqrt{C^2+4C_2}}{C_2}\right)^{\!n}
\Big(C-C\,C_2+\sqrt{C^2+4C_2}+C_2(-4+\sqrt{C^2+4C_2})\Big)
\nonumber\\
&&+
\left(-\,\frac{C-2C_2+\sqrt{C^2+4C_2}}{C_2}\right)^{\!n}
\Big(C(-1+C_2)+\sqrt{C^2+4C_2}+C_2(4+\sqrt{C^2+4C_2})\Big)
\Bigg].\nonumber
     \end{eqnarray}
     The two terms to the power $n$ become in the leading order
     \begin{eqnarray}
         &&2^n\left(1+\frac{1}{C}\right)^n\approx 2+O(n/L^3),\,\,2^{-n}\left(1+\frac{1}{C}\right)^n\approx 2+O(n/L^3),\\
         &&2^{-n}\left(1-\frac{C}{C_2}\right)^n\approx 2e^{-n/L}\approx 0.
     \end{eqnarray}
     Thus, 
     \begin{eqnarray}
         F_n&&\approx \frac{L^3}{(1+C-C_2)^2\sqrt{C^2+4C_2}}\left[C-CC_2+\sqrt{C^2+4C_2}+C_2\left(\sqrt{C^2+4C_2}-4\right)\right]\\
         &&\approx \frac{C_2}{C_2^2 \sqrt{C_2}} \sim \frac{1}{C_2^{3/2}} \sim \frac{L^3}{C^2}+O(L^{-3})+O(e^{-n/L})+O(n/L^3),\nonumber
     \end{eqnarray}
     which, after taking into account the arbitrariness of the initial point, results in the same relative subleading correction.
    Next, we show that the $1/\ell$ scaling regime provides negligible corrections. As the  scaling regime $1/\ell^{5/2}$ emerges for $O(1)\ll n\ll L^2$, its weight scales as $\int_{O(1)}^{L^2}\mathrm d\ell\,L^3\ell^{-5/2}\sim L^3$. At the same time, in the regime $L^3\gg n\gg L^2$, one finds the scale $\int_{L^2}^{L^3}\mathrm d\ell\ell^{-1}\sim \log L$, leading to negligibly small corrections compared to the $1/\ell^{5/2}$ scaling regime. This is also in good agreement with the general consideration that a random walk of length $\ell$ has typical deviations from the origin scaling as $\sqrt\ell$ and so covers a volume of $v_\ell\sim \ell^{3/2}$. The crossover to the $1/\ell$ scaling regime thus takes place when this scale reaches the full volume, $v_\ell\sim L^3\Rightarrow \ell \sim L^2$. At the same time, vortices cannot grow longer than $\ell_\mathrm{max}\sim L^3$, justifying the upper bound for the $1/\ell$ scaling regime.

     Finally, we also show the leading-order scale of the probability that a loop closes at an arbitrary $n$, thereby identifying the probability of loop occurrence. Resorting to long loops, one simply finds that
     \begin{eqnarray}
         \sum_{n=1}^\infty\frac{F_n}{n}\propto \sum_{n=1}^\infty q^{1/2}n^{-5/2}=q^{1/2}\zeta(5/2)\propto q^{1/2}.
     \end{eqnarray}

    \section{First two vortex line length cumulants}
    In this section, we show that the 3$D$ random walk picture immediately implies the non-proportional universal scaling law for the mean and variance of the total vortex strings. The key ingredient is the scaling of the average number of vortex loops, $\kappa_1(N)\sim\hat\xi^{-3}$, and the length of the individual vortex line elements, $\ell_0\sim\hat\xi$ fixed by the nonequilibrium scaling theory, for slow or fast processes. Here, the freeze-out length scale follows the universal power laws
    \begin{eqnarray}
        \hat\xi\propto \tau^{1/4}_Q,\qquad\hat\xi\propto \epsilon^{-1/2}_f,
    \end{eqnarray}
    for slow and fast driving, respectively. To this end, the average is given by the average over the number of line elements in the loops and the number of loops themselves, i.e., the average of the sum $L=\sum_{i=1}^N\ell_0n_i$, with both $n_i$, the number of line elements in the $i$-th loop, and $N$ being random.
    This can conveniently be handled by the tower rule as a powerful technique to evaluate conditional expectation values,
    \begin{eqnarray}
        \langle L\rangle=\ell_0\mathbb E\left[\sum_{i=1}^N n_i\right]=\ell_0\mathbb E\left[\mathbb E\left[\sum_{i=1}^Nn_i\right]\Big\vert N\right]=\ell_0\mathbb E\left[N\mathbb E\left[n_i\right]\right]=\ell_0\langle N\rangle \langle n_i\rangle\propto\hat\xi^{-2}q^{-3/2},
    \end{eqnarray}
    where we have used that the average loop length is independent of $\ell_0$ and inherits the $q$ dependence of the length distribution. Here, the inner mean $\mathbb E[\cdots]$ denotes averaging the number of steps in the individual loops. The outer average $\mathbb E[\cdots]$ runs over the number of loops. The first average was evaluated by the tower rule, i.e., by taking the conditional average $\mathbb E\left[\cdots\Big\vert N\right]$ of the individual loops for a given fixed number of loops $N$.
    
    Assume that loop-loop correlations are homogeneous, i.e., $\mathrm{cov}(n_i,n_j)=c$
    \begin{eqnarray}
        \kappa_2(L)&&=\ell^2_0\mathbb E\left[\sum_{i=1}^N\kappa_2(n_i)+2\sum_{i<j<N}\mathrm{cov}(n_i,n_j)\Big\vert N\right]+\kappa_2(N)\kappa^2_1(n_i)=\ell^2_0\mathbb E\left[\mathbb E\left[\sum_{i=1}^N\kappa_2(n_i)+2\sum_{i<j<N}\mathrm{cov}(n_i,n_j)\right]\Big\vert N\right]\nonumber\\
        &&=\ell^2_0\kappa_1(N)\kappa_2(n_i)+\ell^2_0 \mathbb E\left[N(N-1)c\right]+\ell^2_0\kappa_2(N)\kappa^2_1(n_i).
    \end{eqnarray}
    Assuming homogeneous correlations between the loops, within the picture that the growth of one loop restricts the available space for the others, one can consistently impose $\mathrm{cov}\left(\sum_{i=1}^Nn_i,n_j\right)=0$ for a fixed $N$ number of loops, one finds
    \begin{eqnarray}
        \mathrm{cov}\left(\sum_{i=1}^N n_i,n_j\right)=\mathrm{cov}(n_i,n_i)+2\sum_{i<j<N}\mathrm{cov}(n_i,n_j)=\kappa_2(n_i)+(N-1)c=0\Rightarrow c=\kappa_2(n_i)/(N-1),
    \end{eqnarray}
    from where the second part under the conditional expectation cancels the first part, yielding
    \begin{eqnarray}
        \kappa_2(L)=\ell^2_0\kappa_2(N)\kappa^2_1(n_i)\propto \hat\xi^{-1}q^{-3},
    \end{eqnarray}
    which matches well the predicted non-proportional power law in the main text.

\section{Freeze-out time-scaling for slow and fast drivings}

\begin{figure}[h]
        \includegraphics[width=.75\linewidth,trim={0cm 4cm 0cm 10cm},clip]{ 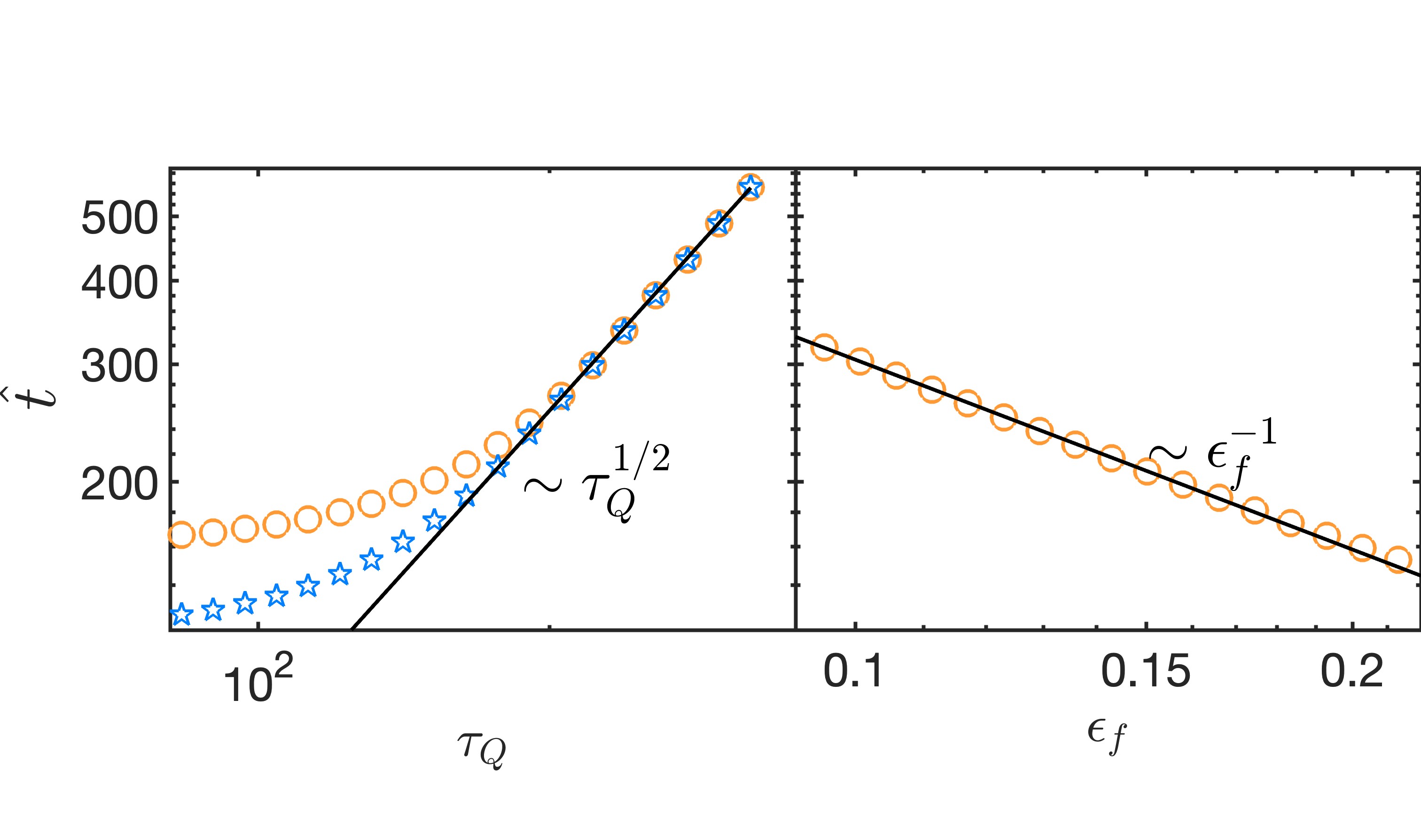}
    \caption{
        (a) Freeze-out time scale for slow drivings with $\hat{t}=(9.09\pm0.43)\tau_Q^{0.483\pm0.006}$ (b) Fast-quench scaling $\hat{t}=(34.6\pm0.6)\epsilon_f^{-0.945\pm0.008}$. Both fast and slow regimes are in close agreement with the theoretical predictions.   
    }
    \label{fig:t_freeze}
\end{figure}

In this section, we provide additional numerical evidence for the universal signatures of slow and fast dynamics in terms of the freeze-out time. For sufficiently large $\tau_Q$, the freeze-out time follows precisely the predicted scaling $\hat t\propto\tau^{1/2}_Q$, while for fast quenches it agrees well with the theoretical power law, $\hat t\propto\epsilon^{-1}_f$, as shown in Fig.~\ref{fig:t_freeze}(a) and Fig.~\ref{fig:t_freeze}(b), respectively.

\section{Precision of the Gaussian fits to the vortex loop number and length histograms}
In this section, we provide further technical details of the validation of the Gaussian fits to the numerically obtained histograms in Fig.~\ref{fig4} and Fig.~\ref{fig:L_totStat}.
We quantify the precision of the Gaussian fit by its normalized trace distance from the numerical results, defined as
 \begin{eqnarray}
    D_\mathrm{TN}=\frac{\frac{1}{2}\sum_n\lvert p^{(\Delta)}_n-q^{(\Delta)}_n\rvert}{\sum_nq^{(\Delta)}_n},
 \end{eqnarray}
 for two arbitrary distributions where the empirical one $q^{(\Delta)}_n$ is constructed using bin sizes $\Delta$ to count the frequency of events $q_n$ in the interval $[(n-1)\delta,n\Delta],\,n=1,2,3,\dots$ For the histograms of the vortex loop number, we used $\Delta=1$, which provides the highest resolution and yields small trace distances. In the fast and slow driving regimes, 
 \begin{eqnarray}
     &&D_{\mathrm{TN},\epsilon_f=0.136}=0.086,\quad D_{\mathrm{TN},\epsilon_f=0.175}=0.073,\\
     &&D_{\mathrm{TN},\tau_Q=244,T_f=0.7\,T_c}=0.1,\quad D_{\mathrm{TN},\tau_Q=665,\,T_f=0.8T_c}=0.093.
 \end{eqnarray}
 For the vortex loop length distributions, we use a bin size $\Delta=100$ in agreement with the typical value of single loop lengths, i.e., matching the scale of $\langle L\rangle/\langle N\rangle\sim 10^2$ for the reported data.
Even a better matching with smaller trace distances is found for the same set of parameters,
\begin{eqnarray}
     &&D_{\mathrm{TN},\epsilon_f=0.136}=0.05,\quad D_{\mathrm{TN},\epsilon_f=0.175}=0.029,\\
     &&D_{\mathrm{TN},\tau_Q=244,T_f=0.7\,T_c}=0.048,\quad D_{\mathrm{TN},\tau_Q=665,\,T_f=0.8T_c}=0.052.
 \end{eqnarray}

\end{appendix}

\end{document}